**EXPLORING A SPECIALIZED ECCENTRICITY-BASED DEEP NEURAL NETWORK MODEL TO SIMULATE VISUAL SEARCH PROCESS**

Major project report submitted in partial fulfillment of the requirements for the award of

Master of Science

in

**Cognitive Science**

by

**Manvi Jain**

2023HCS7008

Under the
Guidance of

**Prof. Sumeet Agarwal, Prof. Pawan Sinha, Dr. Priti Gupta**

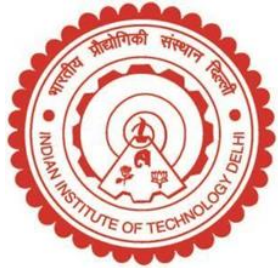

**DEPARTMENT OF HUMANITIES AND SOCIAL SCIENCES**

**INDIAN INSTITUTE OF TECHNOLOGY DELHI**

**NEW DELHI - 110016**

**MAY 2025**

# UNDERTAKING BY THE STUDENT

I hereby declare that the work presented here in the report has been carried out by me towards the partial fulfillment of the requirement for the award of Master of Science in Cognitive Science at the Department of Humanities and Social Sciences, Indian Institute of Technology Delhi. The content of this report, in full or in parts, have not been submitted to any other institute or university for the award of any degree.

**Student Name: Manvi Jain**

**Entry number: 2023HCS7008**

**Contact No.:**

**Email Id.: hcs237008@iitd.ac.in**

**Place: Delhi**

**Date: 5th May 2025**

# CERTIFICATE BY THE SUPERVISOR

This is to certify that the report entitled, "EXPLORING A SPECIALIZED ECCENTRICITY-BASED DEEP NEURAL NETWORK MODEL TO SIMULATE VISUAL SEARCH PROCESS" being submitted by Manvi Jain (2023HCS7008) to the Department Humanities and Social Sciences, Indian Institute of Technology Delhi for partial fulfillment of the requirement for the award of degree of Master of Science in Cognitive Science. This study was carried out by her under my guidance and supervision.

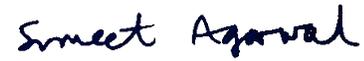

**Signature of the supervisor(s)**

**Prof. Sumeet Agarwal**

**Yardi School of Artificial Intelligence,**

**Indian Institute of Technology Delhi**

**Place: Delhi**

**Date: 5th May 2025**

# ACKNOWLEDGEMENTS

I am grateful for the kind supervision and continued support of all my supervisors throughout the project. I also acknowledge the support of my friend Shishank Jain in terms of resources required to run the models throughout the duration of this project.

# ABSTRACT

There exist many theories and models of visual attention for humans and animals alike, however, a very small area of research is able to contribute to our understanding of the neurodevelopmental trajectory of visual attention and related processes. This is owed to the fact that during their actual development, humans are not very reliable for experimentation, as in the infancy stage, individuals tend to have shorter attention span towards a specific stimuli, making their responses unreliable and hard to gather. Therefore, we collaborate with project Prakash which provides a unique window into the world of neurodevelopmental studies with patients who have gained vision later in their life, thus, beginning to develop the various sophisticated visual processes at a later stage than infancy where they are more responsive and reliable. This unique opportunity provided us with the data from human participants coming from different groups such as patients pre-operation and post-operation, several months after operation at regular intervals (one-month, three-months, six-months and one-year), neurotypical subjects (or controls) wearing blurred goggles to match the mean visual acuity of the patient group after sight-restoration surgery. The data consists of participants' responses to a modified pre-attentive pop-out visual search task. In an effort to model the cognitive processes involved in evaluating these kinds of tasks, we explored and experimented with a guided-search theory based CNN model called eccNET model designed by Gupta et al., (2021) to simulate visual search asymmetry computationally. The model was given a similar task as the human participants to perform visual search. The qualitied comparison of their RTs provided insights into the similarities and differences between the human and model performance, thus, providing a unique opportunity to try model ablations to adjust its implementational processes based on the known facts about the human physiology involved in the process and correspondingly with the patient group at different times in their developmental journey. The differences were more important as they allowed us to focus on developing ideas around the enhancements required in the model to make it appear more like humans in their developmental journey.

# CONTENTS

# LIST OF FIGURES

# CHAPTER 1: INTRODUCTION

## 1.1 Visual Attention and Visual Search

Visual attention is the mechanism that the nervous system uses to highlight specific locations, objects or features within the visual field. This can be accomplished by making an eye movement to bring the object onto the fovea (overt attention) or by increased processing of visual information in neurons representing more peripheral regions of the visual field (covert attention) (Bisley, 2011). Searching involves directing attention to objects that might be the target (overt attention). This deployment of attention is not random. It is guided to the most promising items and locations by five factors discussed here: bottom-up salience, top-down feature guidance, scene structure and meaning, the previous history of search over timescales ranging from milliseconds to years, and the relative value of the targets and distractors (Wolfe & Horowitz, 2017). The process extensively deploys eye movement determined mostly by stimulus salience, object recognition, actions, and value (Schutz et al., 2011).

## 1.2 Theories of Visual Search

### *1.2.1 Feature Integration Theory (FIT)*

Treisman & Gelade (1980) established a hierarchical framework for understanding visual search processes that occurs in two main stages: Simple and Conjunction Search. According to FIT (Feature Integration Theory), the process of visual search begins with the detection of local features in an image, where specific neurons respond to these features and form a feature map. Different features are then highlighted in a location map, indicating where the searched feature is located. Simple search processing begins with the processing of visual elements and neural responses to local features, followed by the generation of feature maps and location mapping activation. Conjunction search processing involves serial attention deployment, feature-based filtering, object-memory integration, and the management of illusory conjunctions.

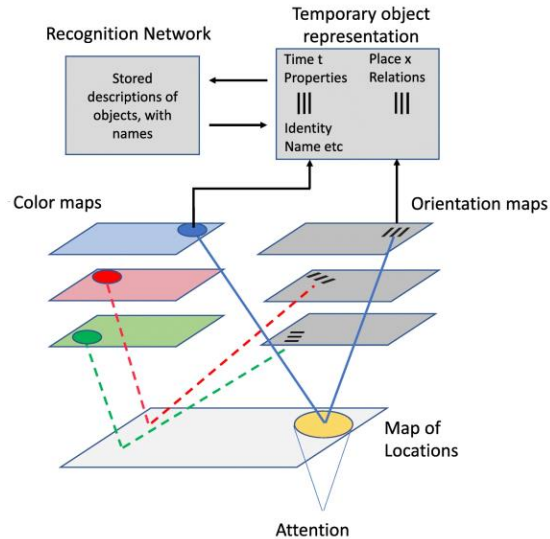

Figure 1: The theoretical framework of the Feature Integration Theory (FIT). A particular stimulus creates activations in feature maps (for color and orientation in this case). Attention then binds features together in the master map of locations, but can only do so for a limited amount of information in the display. The temporary object representation is then compared against stored object descriptions. (Based on Fig. 1 in A. Treisman & Gormican, 1988).

*Simple Search Processing (Feature Search)* represents the initial, parallel stage of visual processing where basic features are automatically extracted and processed across the entire visual field. When visual information first enters the system, specialized neural populations in different areas of the visual cortex simultaneously process fundamental properties such as color, brightness, orientation, motion, and size. In V1, neurons respond to basic edge detection and orientation, while V2 processes simple shapes, V4 handles color and form, and MT/V5 processes motion information. This distributed neural response leads to the generation of distinct feature maps - topographic representations where each map corresponds to a specific feature dimension like color or orientation. These maps maintain spatial information while showing the presence, intensity, and distribution of features across the visual field. The process culminates in location mapping activation, where spatial information is integrated to create a master map that preserves the locations of various features.

*Conjunction search processing* represents a more complex, serial stage of visual processing that requires focused attention and conscious control. During this stage, attention is deployed serially across the visual field, moving from location to location to bind different features together. The process employs feature-based filtering, where attention selectively enhances relevant feature combinations while suppressing irrelevant ones. This filtering mechanism helps manage the overwhelming amount of visual information by prioritizing processing based on task demands. The system then integrates these bound features with stored object memories, comparing current percepts against known object representations to achieve recognition. Throughout this process, the visual system must actively manage illusory conjunctions - incorrect feature combinations that can occur when attention is overloaded or divided. This management involves mechanisms for error detection and correction, ensuring accurate object perception despite the complexity of natural scenes and the limited capacity of attention.

This two-stage model explains why searching for a single feature (like a red item among green ones) is typically fast and effortless while searching for a conjunction of features (like a red circle among red squares and green circles or multicolored circles) requires serial processing and is consequently slower and more demanding. The model also accounts for common visual search phenomena such as pop-out effects, search asymmetries, and the role of distractor heterogeneity in search efficiency.

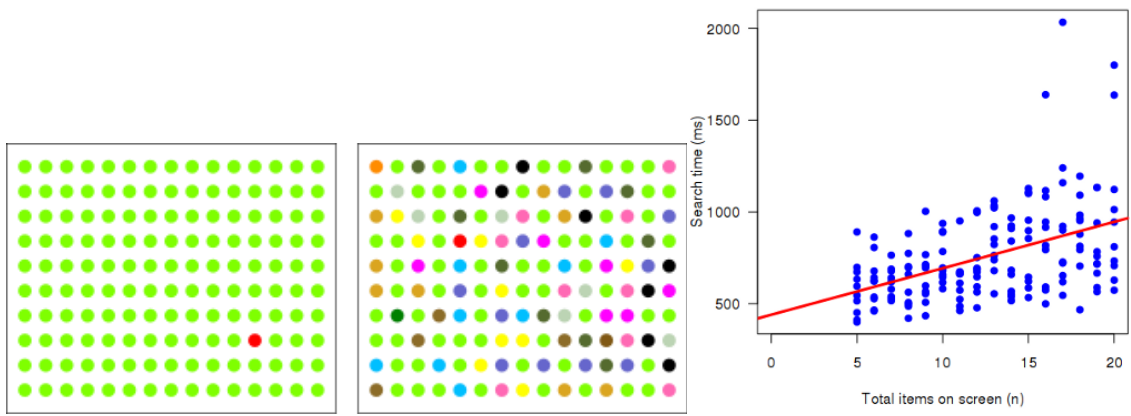

Figure 2: Simple search and conjunction search tasks represented with the reaction time corresponding to increase in number of stimuli on screen referred to as set size. *Source*: psytoolkit.com

*1.2.2 Guided Search Theory (GST)*

Wolfe (1994) expanded our understanding of visual search by proposing an integrated approach that emphasizes the interaction between bottom-up and top-down processes. This theory describes how feature maps are preactivated based on expectations, and how visual processing proceeds serially from the most to least salient features. GST posits that visual search is not solely driven by bottom-up processes, such as the detection of these visual elements, but is also influenced by top-down cues and biases. In this view, pre-existing expectations or knowledge about the scene play an important role in guiding attention and refining the search process. Thus, both bottom-up and top-down factors interact to shape visual perception and search behavior.

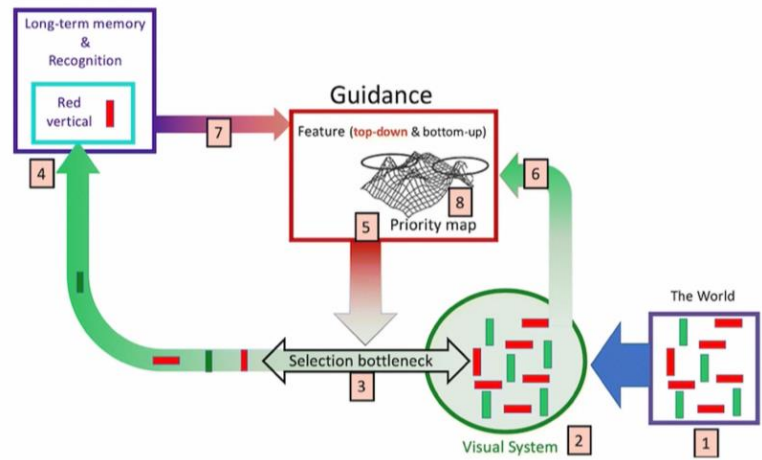

Figure 3: A schematic representation of Guided Search 2.0. (Wolfe et al., 2021)

Visual search is significantly influenced by action-related processes and competitive mechanisms. Top-down guidance integrates visual properties with action information, while accounting for the impact of eye movements on stimulus processing and action-stimulus congruency effects (Bekkering & Neggers, 2002). Action-based attention enhances processing for action-relevant objects and creates location-specific advantages. The Biased Competition Theory (BCT) provides a framework for understanding how multiple attention processes are integrated, how selection occurs based on attentional weights, and how coherent behavior emerges through bias integration (Desimone & Duncan, 1995).

*1.2.3 Guided Search Model 6.0*

Treisman's feature integration theory is one of the earliest theories that gave a broad view of how visual search is carried out. It got an addition in terms of the possibility that visual search is driven by bottom-up feature of stimulus in Wolfe's guided search model. Guided search 6.0 model of visual search is one of the latest ones that give an overview of the same (Wolfe, 2021). The following is an account of the model: GS6 is an advanced model of visual search that describes how humans efficiently locate targets in cluttered scenes despite limited attentional capacity. It posits that selective attention is guided by five preattentive sources of information, integrated into a dynamic priority map that directs eye movements approximately 20 times per second:

1. Top-down feature guidance : Knowledge of target features (e.g., "look for red") biases attention toward relevant items.

2. Bottom-up feature guidance: Salient differences (e.g., a bright spot in a dim field) draw attention automatically.

3. Prior history: Recent experiences, like priming, influence what's prioritized (e.g., repeatedly searching for circles boosts circle detection).

4. Reward: Items associated with value gain priority.

5. Scene syntax and semantics : Contextual understanding (e.g., cups are on tables) guides attention to likely locations.

The priority map evolves over time, favoring items near fixation due to foveal biases described by three functional visual fields (FVFs): resolution (sharpness decreases with eccentricity), selection (attention prioritizes near-fixation items), and binding (feature integration is stronger centrally). In typical adults, this system efficiently narrows search by combining preattentive cues with attentional selection.

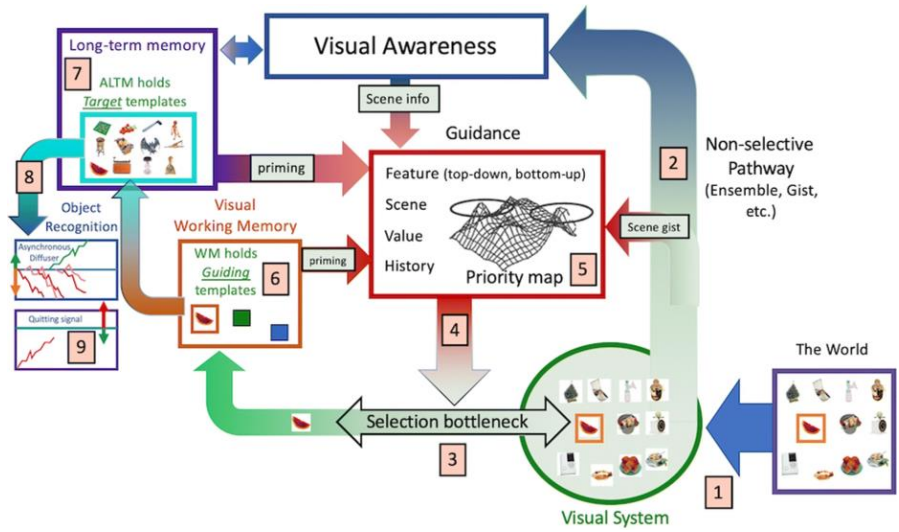

Figure 4: A schematic representation of Guided Search 6.0

# CHAPTER 2: LITERATURE REVIEW

## 2.1 Computational Models of Visual Search

### 2.1.1 Search by Recursive Rejection Model

The Search by Recursive Rejection (SERR) model, developed by Humphreys & Müller (2020), is an early connectionist model designed to explain how features are processed and targets are selected among distractors during visual search (Figure 5). This model employs computationally simple processing units that sum activation from input units and pass it on to other units involved in response selection. A key advantage of such connectionist models is that activation within the network evolves dynamically over time, making them particularly useful for simulating the time course of information processing. In SERR, individual feature maps (e.g., for horizontal and vertical edges) are initially encoded, and activation is subsequently passed to conjunction units that represent relationships between these features, such as specific corner configurations at different orientations. The model incorporates mutual excitation between similar stimuli across different locations, which amplifies the activation of targets, and inhibitory interactions between non-matching conjunctions at the same location. This dynamic results in a super-additive effect, whereby the presence of two targets facilitates faster detection than would be expected from the summed performance of detecting each target alone, aligning well with empirical human data. Once activation reaches the level of target or non-target templates, decisions are made: if the target template surpasses a threshold, a "present" response is triggered, whereas activation of a non-target template leads to its inhibition and recursive suppression of associated conjunction units. This rejection process continues until the target is identified or all distractors are eliminated. Despite its strengths in modeling critical aspects of grouping in search, SERR has notable limitations. The model relies on hardwired grouping relations rather than adaptive or learned ones, lacks mechanisms for top-down modulation based on expectations or task goals, and does not extend to higher-level object recognition, as its conjunctive units are not organized into coherent object representations.

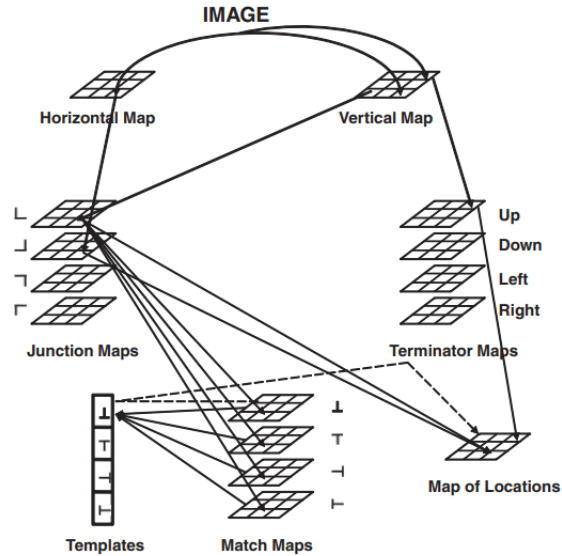

Figure 5: A schematic representation of working of SERR Model

*2.1.2 Selective Attention for Identification Model*

The Selective Attention for Identification Model (SAIM), proposed by Heinke & Humphreys, (2003), was developed to account for both the selection of a target among distractors and the recognition of objects in a translation-invariant manner (Figure 6). Unlike earlier models, SAIM introduces a dual-network architecture, where each pixel in the visual input is simultaneously mapped onto two parallel systems: the content network and the selection network. The content network encodes form information derived from the retinal image, while the selection network facilitates attentional focus by determining which portions of the content are passed to a "focus of attention" (FOA). The FOA consists of units that correspond to stored object templates within a knowledge network. Within the selection network, units engage in competitive interactions—mutual inhibition occurs when multiple mappings target the same FOA unit, whereas mutual excitation supports mappings to different FOA units. This dynamic enables the system to isolate and enhance relevant portions of the visual field for identification. Additionally, SAIM incorporates top-down modulation through input from the knowledge network, which influences the competition within the selection network based on prior knowledge or expectations. Despite its sophisticated architecture, SAIM is limited in several ways. It operates at the pixel level, making it less effective at grouping visual elements or generalizing to novel forms. Its reliance on stored templates for grouping further restricts its flexibility in handling unfamiliar stimuli. Moreover, the use of simple processing units, while computationally efficient, lacks the biological realism needed for direct comparisons with neural data, thus limiting its applicability in modeling physiological processes underlying visual attention and recognition.

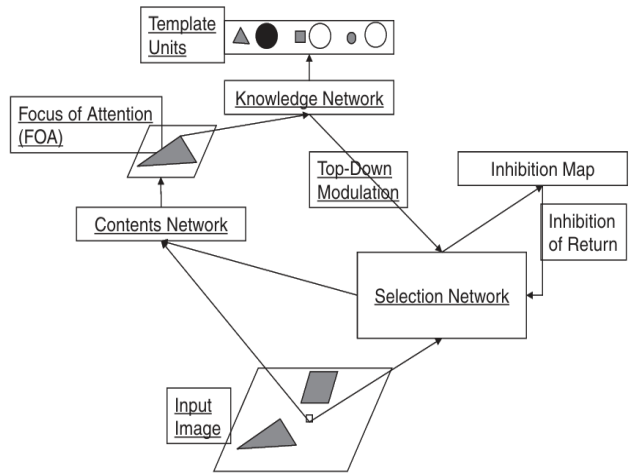

Figure 6: A schematic representation of working of SAIM Model

*2.1.3 Visual Search Implementation of the Selective Attention for Identification Model*

The Visual Search implementation of the Selective Attention for Identification Model (VS-SAIM) (Heinke & Backhaus, 2011) was developed as an extension of the original SAIM framework to better address complexities observed in visual search behavior (Figure 7). While SAIM used rate-coded neuron models to simplify neurophysiological processes and was effective in simulating attentional selection and visual search, it faced limitations when applied to a broader range of empirical findings. To improve its performance, VS-SAIM introduced a more refined similarity measure, replacing the original with a Euclidean distance metric, which allowed for more accurate modeling of perceptual similarity between stimuli. This modification enabled VS-SAIM to capture challenging phenomena such as search asymmetry—where the ease of detecting a target among distractors is not reciprocal. The core functioning of VS-SAIM remains consistent with SAIM's principles, achieving object recognition through translation-invariant mapping from retinal input to a focus of attention (FOA). Within this process, competitive interactions among units ensure the selection of the most relevant object representations, while cooperative mechanisms help maintain spatial neighborhood relationships, preserving the structural coherence of objects. Additionally, top-down pathways provide a bias toward known or task-relevant objects, enhancing the model's attentional selectivity. Despite these advancements, VS-SAIM still has notable limitations. Its neural architecture, while inspired by biological systems, lacks full biological plausibility, restricting its application in accurately modeling complex brain functions. Furthermore, the model's dependence on template matching and Euclidean distance metrics oversimplifies the variability inherent in real-world object recognition, making it less effective for novel or variable stimuli. Scalability also remains a concern, as the computational demands of VS-SAIM may limit its efficiency, particularly in processing large-scale or dynamic visual scenes in real-time applications.

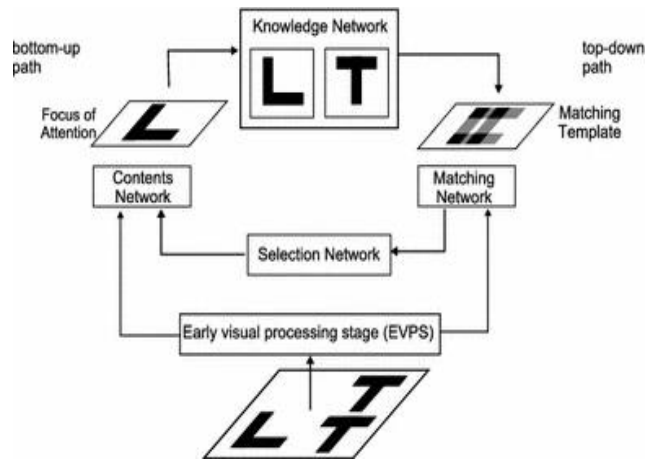

Figure 7: A schematic representation of working of VS-SAIM Model

*2.1.4 eccNET Model*

The proposed model builds upon foundational theories in visual attention, particularly feature integration theory, the guided search model, and the invariant visual search network (Gupta et al., 2021). Central to this model are two primary mechanisms of attentional guidance: bottom-up, stimulus-driven saliency, and top-down, target feature-based modulation. Unlike earlier models that often focused on one of these mechanisms in isolation, this model integrates both, offering a more comprehensive account of visual attention. To achieve this, it utilizes a biologically inspired representation of the human ventral visual cortex through a pre-trained deep convolutional neural network (CNN), incorporating eccentricity-dependent sampling to mimic neurophysiological constraints. This design allows the model to extract spatially sensitive visual features from both the target and the search image.

The model operates in a sequence of fixations, simulating the human process of visual search. It takes two inputs: a target image, representing the object to be found, and a search image, containing the target embedded among distractors. Starting with a central fixation on the search image, the model computes at each fixation a bottom-up saliency map and a top-down attention map. These are linearly combined to produce an overall attention map, and a winner-take-all strategy is employed to select the next fixation point. The model continues this iterative process until the target is located, with the assumption of infinite inhibition of return, meaning previously fixated locations are never revisited. Although this differs from human search behavior—where return fixations are possible—it is justified here by the limited number of fixations typically required in the experiments considered. Refer to Figure 8 for schematic observation of the process. Target verification at each fixation is handled by an "oracle" recognition system, which checks whether the fixation lies within the ground truth bounding box of the target object. This bypasses complex recognition steps to focus on the search mechanism.

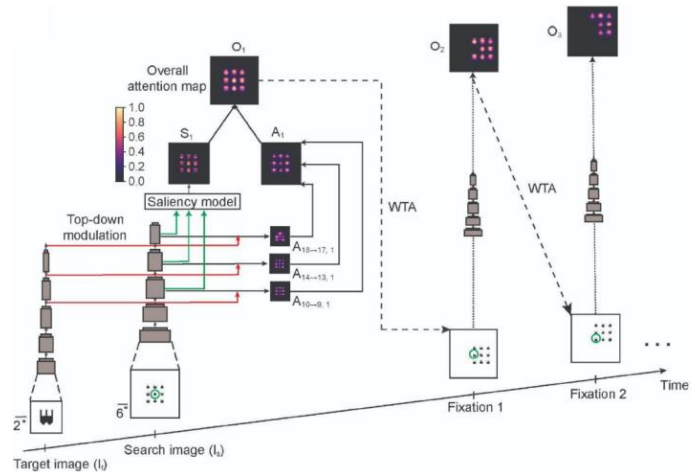

Figure 8: A schematic representation of eccNET model

*2.1.4.1 Visual representation at the neural level and computational level*

Artificial Neural Networks (ANNs) are fundamentally inspired by the structure and function of biological neurons. While artificial neurons operate using continuous real-valued signals, biological neurons communicate via discrete electrical impulses or spikes. However, much of the information in biological systems is conveyed through the frequency of these spikes rather than their precise timing. This allows for a meaningful analogy between spike frequency and the real-valued outputs of artificial neurons, making it reasonable to consider artificial neurons as simplified models of their biological counterparts. Convolutional Neural Networks (CNNs), a subclass of ANNs featuring convolutional layers, have been particularly influential in modeling visual processes. They provide a strong computational analogy to the simple and complex cells observed in the V1 area of the ventral visual stream, as early CNN layers have been shown to respond to basic visual features such as edges and colors, much like the V1 cortex (Figure 8).

Integration of Computational and Neural Insights in the model: The eccNET framework demonstrates that search asymmetry emerges spontaneously in systems trained on natural images, without task-specific fine-tuning. This supports GS6's premise that search efficiency arises from: Architectural constraints (e.g., foveal bias) and Statistical regularities in visual experience. For instance, eccNET's difficulty in detecting straight lines among curves mirrors human behavior, as curved edges are overrepresented in natural scenes. Such findings validate deep networks as tools for reverse-engineering perceptual biases rooted in both biology and environment.

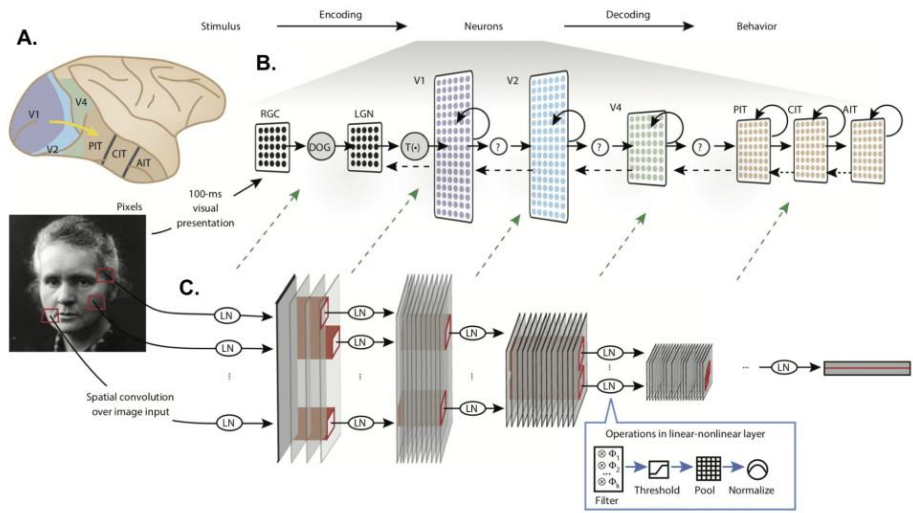

Figure 8: Structure of visual pathway in the brain and machine, A. shows the anatomical brain regions, B. shows the model ventral visual stream, C. shows an arbitrary convolutional neural network architectures (Yang & Wang, 2020).

Functional similarities between Deep CNNs and the ventral stream that are used in the current model to simulate human-like physiological structure in visual attention process:

| **Biological System** | **Computational Model** |
|---|---|
| V1 simple/complex cells | Early CNN edge/orientation filters |
| Ventral hierarchy (V1 → IT) | Deep CNN layer hierarchy |
| Eccentricity-dependent RFs | eccNET's adaptive pooling layers |
| Top-down attention | Target-guided feature modulation |

*2.1.4.2 Neurophysiological Resemblance in eccNET*

eccNET closely mirrors several key neurophysiological properties of the primate visual system:

1. Eccentricity-dependent receptive fields: The model's pooling layers mimic the increase in receptive field size with retinal eccentricity observed in primate visual cortex (e.g., V1, V2, V4, IT). This is critical for reproducing the foveated nature of primate vision, where fine detail is processed centrally and coarse information peripherally (Figure 9).
2. Hierarchical processing: Like the ventral stream, eccNET processes visual information through a sequence of layers, each integrating features from earlier stages to form progressively complex representations-from simple edges and orientations in early layers to complex object features in deeper layers.
3. Top-down attentional modulation: The model's mechanism for modulating activations based on target features parallels the influence of fronto-parietal circuits on ventral stream activity during visual search, implementing a biologically plausible form of selective attention.

4. Dynamic fixation selection: eccNET's winner-take-all mechanism for choosing successive fixation points simulates saccadic eye movements driven by a priority map, consistent with neurophysiological evidence of attention-guided eye movements in primates.

While eccNET is not a perfect quantitative model of primate neurophysiology, it preserves essential trends such as receptive field scaling with eccentricity and hierarchical feature integration, providing a computational framework that bridges neural data and behavioral phenomena.

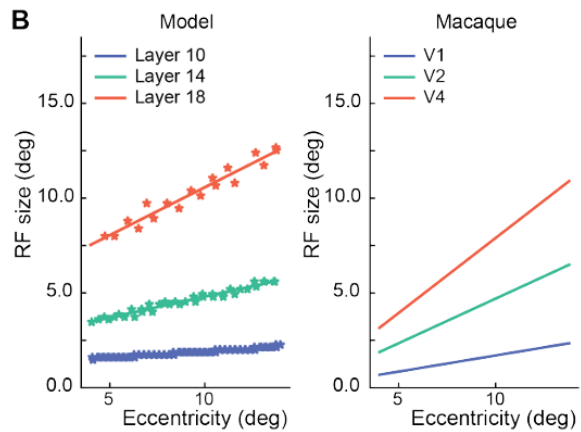

Figure 9: Eccentricity-dependent sampling leads to increasing receptive field sizes as a function of eccentricity, in addition to increased receptive field sizes across layers for the model (left) and also for the macaque visual cortex

*2.1.4.3 Standardization Against Similar Related Models*

eccNET builds upon and advances prior computational models of visual search, notably the Invariant Visual Search Network (IVSN) and classical CNN architectures:

1. Compared to standard CNNs (e.g., VGG16): Traditional CNNs assume uniform spatial resolution and fixed pooling sizes across the visual field. eccNET innovates by replacing standard max-pooling with eccentricity-dependent pooling layers, introducing spatially varying receptive field sizes that better reflect biological vision.
2. Compared to IVSN: While IVSN uses a static attention map and applies top-down modulation at a single layer, eccNET dynamically updates the attention map at each fixation and integrates top-down modulation across multiple layers. This allows eccNET to more accurately capture the dynamic nature of visual search and attentional shifts.

The eccNET model was also tested against the following four baseline models. eccNET consistently outperformed these models across various visual search tasks, demonstrating that its biologically-inspired architecture better captures human visual search patterns. The performance gap was particularly evident in complex scenes and when searching for targets that required integration of features across multiple levels of visual processing:

1. Chance model: This simple baseline generates eye fixations through uniform random sampling, providing a bottom-floor comparison for any intelligent search system.
2. pixelMatching: This approach generates an attention map by sliding the raw pixels of the target image over the search image with a stride of 1×1, representing a basic template-matching approach.
3. Graph-Based Visual Saliency (GBVS): This model uses bottom-up saliency to create the attention map, focusing on inherent visual properties that naturally draw attention regardless of the search target.
4. IVSN: This model creates a top-down attention map based on features from only the top layer of VGG16, representing a more sophisticated but still limited approach.

In summary, eccNET represents a standardized, biologically inspired computational model that integrates neurophysiological principles and deep learning advances to explain complex visual search behaviors, setting a new benchmark for models bridging neuroscience and artificial intelligence.

**2.2 Research Objectives**

This thesis project comprises two principal components. The first is devoted to empirical investigation, wherein we designed and implemented visual search paradigms across diverse participant cohorts. The experimental protocol, detailed in the methodology section, draws upon established frameworks from Wolfe (1994). Participant groups encompassed: normative controls with unimpaired vision (normal controls), controls with artificially degraded acuity (blurred controls), and congenitally blind individuals with treatable bilateral cataracts assessed at multiple postoperative intervals (immediate postoperative, one month, six months, one year, and beyond one year). The data for this special population was sourced through Project Prakash (refer methodology section for more details). Our primary objective was to systematically document performance metrics across these cohorts to elucidate the developmental trajectory of visual search capabilities following sight restoration.

The second component focuses on computational modeling of visual search processes. The foundational architecture derives from Gupta et al. (2021), which operationalizes Wolfe's (1994) Guided Search theory within a computational framework. Our objectives here are twofold: first, to validate the model's performance against our empirical data from human participants; and second, to refine the model's structural and physiological parameters based on current neurophysiological understanding of visual development. These refinements aim to simulate performance patterns observed at various stages of neurodevelopmental recovery following sight restoration surgery, thereby providing mechanistic insights into the neural substrates of visual search acquisition.

# CHAPTER 3: METHODOLOGY

## 3.1 Experimental Studies

### *3.1.1 Visual search task with working memory component requirement*

In this experiment, the participants were required to search for a target bar tilted at an angle of 60° among a specific set of distractors bars of different orientations. Each bar was 4.41 cm long irrespective of orientation and set size. The bars were spread across a white screen of 16:9 aspect ratio designed in Microsoft powerpoint 365. The two variables of interest were set size (5, 10 and 15) and angles (15, 30 and 45 degrees in both negative and positive orientations). Amongst all the trials, 75% were the target-present trials and 25% were the target-absent trials. On target trials, the target item replaced one of the distractor items, so the set size remained constant. Set size, positions of target and distractors, and the presence of a target were randomized across trials. The distribution of trials were kept constant for each category across participants. The distribution of trials in target-present category was as follows: 10 trials for set size 5, set size 10 and set size 15 respectively; target-absent category had a similar distribution except there were no trials for set size 5 in order to avoid increased number of trials. The angles were largely randomly distributed, with more trials having 45° and 30° distractor bars and comparatively fewer 15° bars.

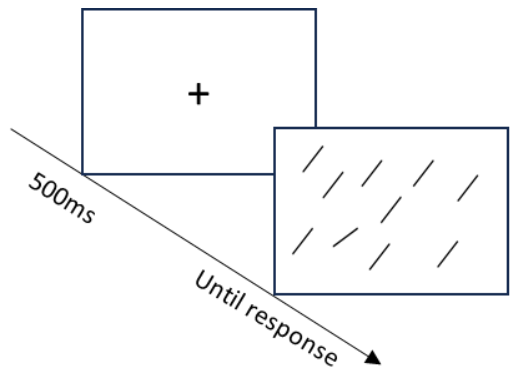

Figure 10: Visual search task paradigm with working memory component

As shown in Figure 10, each stimuli was preceded by a fixation cross (a black plus sign on a white screen) which appeared for 500 msec. The stimuli were presented infinitely until the response was made by the experimenter. The response buttons were allotted as follows: number key 1 was pressed when the participant identified the correct bar, number key was pressed when the wrong bar was identified and space bar was pressed when no identification was made. Reaction time (RT) was calculated for each stimulus. Each control participant was required to wear a blur goggle with acuity 20/500 during the experiment, after which their adapted acuity was measured to make sure they lie in this range. This task was performed by

*3.1.2 Visual search task without a working memory component requirement*

In this experiment, participants were required to search for a target image (displayed on the left side of the screen) among the distractor images enclosed by an outline on the right side. The primary variables of interest were set size (3, 6, 9, and 12).

In 60% of the trials, the target image was present among the distractors (target-present trials), and in the remaining 40%, it was absent (target-absent trials). The trial distribution was held constant across participants: 7 trials for each set size in the target-present condition, and 5 trials in the target-absent condition.

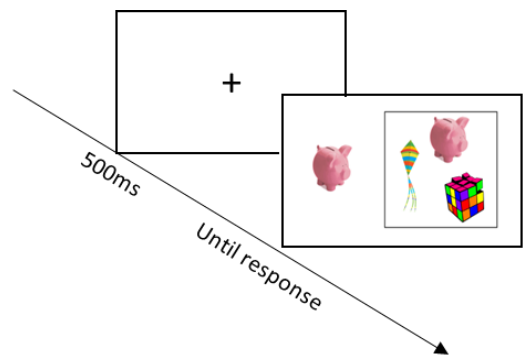

Figure 11: Visual search task paradigm without working memory component

As shown in Figure 11, each trial was preceded by a central fixation cross (a black plus sign on a white screen) presented for 500 milliseconds. The search display remained on screen until a response was made by the participant. Reaction time (RT) was recorded for each trial. To simulate reduced visual acuity, control participants wore blur goggles calibrated to 20/500 vision, and their adapted acuity was measured and reported individually in the results section.

*3.1.3 Data Collection*

The data for human participants for the treatment group was collected at Dr. Shroff's Charity Eye Hospital, New Delhi, India where the philanthropic-research organization, Project Prakash (Sinha, 2016) treats children with congenital cataracts to provide them sight. The ethical clearance for the data collection is sanctioned by the Massachusetts Institute of Technology, USA. The data is collected at different timepoints through the patients' operative journey to understand the progression (or regression) of cognitive capabilities from sight-onset. The data for normal controls was collected at IIT Delhi campus as an online task designed on Psychopy (PsychoPy v2025.1.0).

**3.2 Computational Modeling**

*3.2.1 Theoretical Basis of the eccNET Model*

The eccNET model is grounded in the theory that visual search behavior emerges from the interaction between the hierarchical processing of visual information and attentional mechanisms shaped by both bottom-up sensory input and top-down task demands. The model integrates two core theoretical concepts which are new as compared to the previous models as discussed in literature review section:

5. Eccentricity-dependent visual processing: Inspired by the primate visual system, eccNET incorporates the principle that visual acuity decreases with retinal eccentricity, i.e., resolution is highest at the fovea and progressively lower in the periphery. Unlike standard convolutional neural networks (CNNs) that assume uniform spatial resolution, eccNET introduces *eccentricity-dependent pooling layers* where receptive field sizes increase linearly with distance from the current fixation point. This reflects the biological organization of the ventral visual stream, where receptive fields expand with eccentricity and along the cortical hierarchy (from V1 to IT).

6. **Top-down target-dependent attention:** Visual search is guided by an internal representation of the target that modulates processing of the search image. eccNET extracts features from the target image and uses these to generate a dynamic *attention map* that biases the network's activations toward regions likely containing the target. This top-down modulation is applied across multiple layers of the network, reflecting the distributed nature of attentional influences in the brain.

These principles work together to simulate realistic eye movements, with the model updating where to look based on both the scene layout and target features until it finds what it's searching for. The model's ability to show human-like search patterns without special training supports our theory that these biases naturally emerge from how visual systems process typical images.

### 3.2.2 Model Working Mechanism

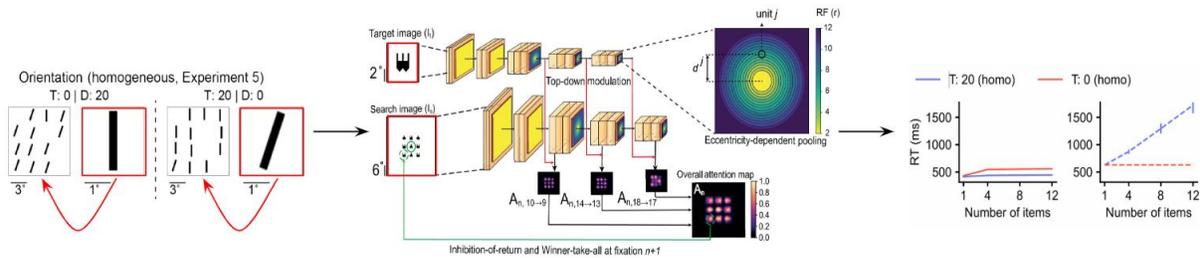

Figure 12: The model inputs a target image and a search image, both of which are processed through the same 2D-CNN model with shared weights. At each fixation n, the model produces a top-down modulation map that directs the next eye movement. The color bar on the right denotes attention values. The corresponding output is shown on the right side.

The eccNET model employs a biologically inspired search mechanism that combines *eccentricity-dependent visual processing* and *target-guided top-down attention* to simulate human-like visual search behavior (Figure 12). The eccNET model works like human vision: it looks at different parts of a scene in succession (eye movements), processing each area with varying clarity based on distance from the current focus point. Built on a modified VGG16 architecture, it processes images with higher detail at the center of "gaze" and coarser detail at the edges, just like our eyes do. At the same time, it uses the target's features to create an attention map that highlights relevant areas, combining automatic salience with purposeful search. The model picks where to look next by choosing the most promising spot on this attention map, continuing until it finds the target. Remarkably, this process works without specific training for search tasks – the search behaviors emerge naturally from features learned through general image recognition and the model's visual structure. This approach bridges what we know about brain function (how the ventral visual pathway works) with computational principles (how deep neural networks process information).

*3.2.3 Conversion of Model Fixations to Reaction Time Measures*

The computational model of visual search employed in this study generates predictions in the form of sequential eye-fixations. However, psychophysical experiments in their study quantified performance through key press reaction times (RT) rather than direct eye movement measurements. This methodological discrepancy necessitated the development of a conversion framework to enable meaningful comparisons between model predictions and human performance data. The underlying assumption guiding this conversion posits that key press reaction time comprises two primary components: (1) cumulative duration of visual processing across multiple fixations, and (2) a constant motor response time required for executing the key press action once the target has been identified.

This relationship can be formalized through the following linear equation:

$$RT = \alpha \times N + \beta \quad (3.11)$$

Where:

- RT represents reaction time measured in milliseconds

- N denotes the number of fixations required to locate the target

- $\alpha$ corresponds to the temporal duration of a single saccade plus the associated fixation period (constant)

- $\beta$ represents the motor response time (constant)

Parameter Estimation: The values for constants α and β were derived using linear least-squares regression analysis applied to the experimental data depicted in Supplementary Figure 5. This analysis yielded the following parameter estimates:

α = 252.36 milliseconds (duration per fixation cycle)

β = 376.27 milliseconds (motor response time)

The resulting model demonstrated robust predictive validity, achieving a correlation coefficient of 0.95 ($p < 0.001$) between predicted and observed reaction times. A critical assumption underlying this conversion framework is that the parameters α and β remain invariant across experimental conditions. Consequently, we applied these constant values uniformly when generating reaction time predictions across all experimental scenarios presented in the various figures. This approach allows for standardized comparisons between model predictions and human performance data, facilitating evaluation of the model's capacity to account for observed patterns in visual search behavior across diverse stimulus conditions.

# CHAPTER 4: RESULTS AND DISCUSSION

## 4.1 Human Participants Performance

The experimental paradigm was structured such that participants indicated target identification by manual gesture, specifically pointing to the target location on the display. The experimenter subsequently evaluated response accuracy in real-time and recorded the judgment via designated key commands. Consequently, reaction time (RT) measurements obtained under these conditions cannot be considered a reliable performance metric due to the introduction of intermediary processing delays. However, as computational model evaluations were quantified using RT metrics, these measurements from human participants were retained for comparative analysis purposes. It should be noted that human performance was independently assessed using accuracy metrics (detailed in the subsequent section) to facilitate exploration of potential confounding variables. The figures presented below illustrate comparative performance metrics across participant cohorts (control subjects and prakash patients) for the experimental tasks described in the methodology section.

### *4.1.1 Reaction Time Performance for Human Participants*

1. Control population & Blurred acuity control population

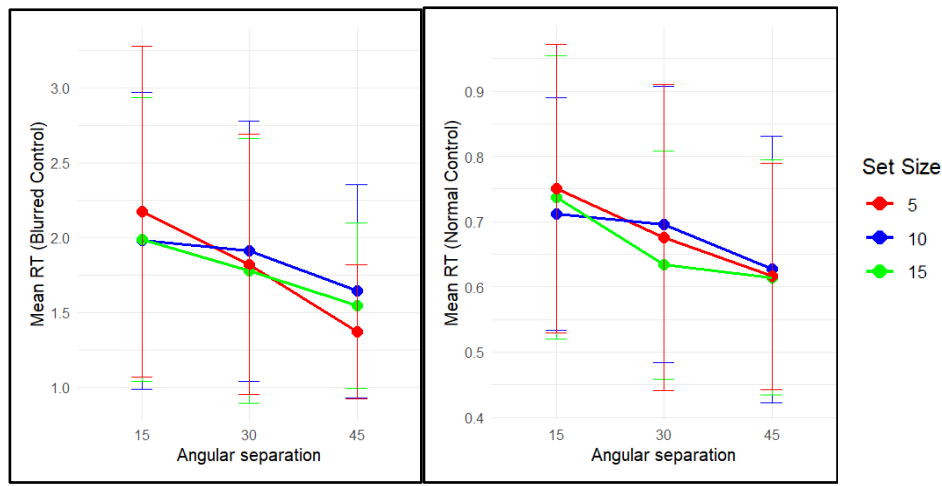

2. Pre-operative and post-operative patient population

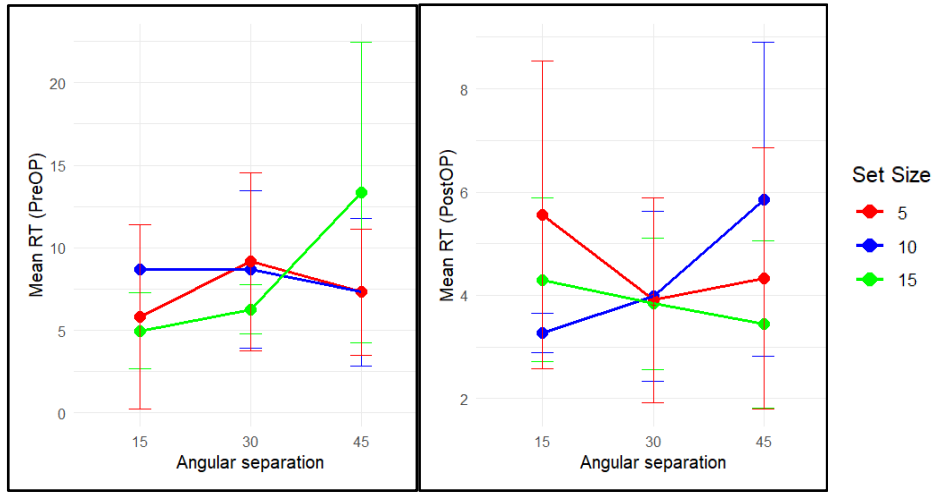

3. Intermediate periods in the post-operative journey of the patient population

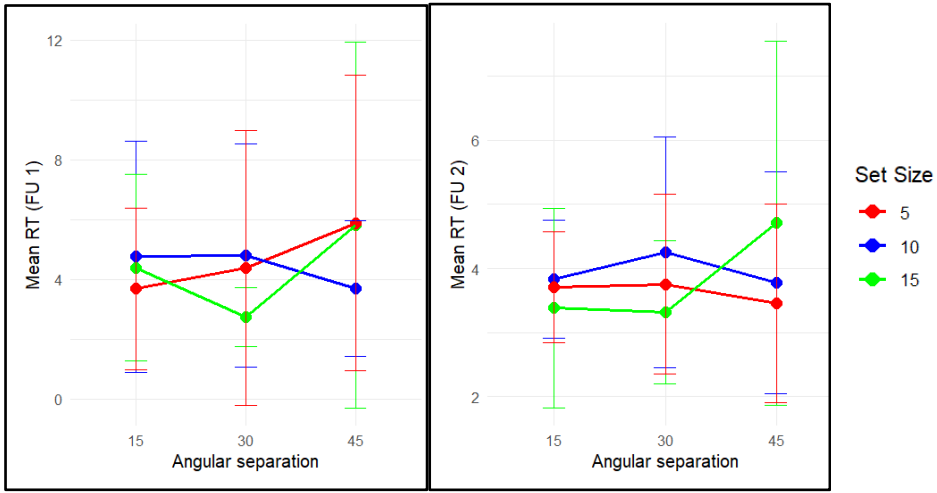

4. Patients treated for more than one year

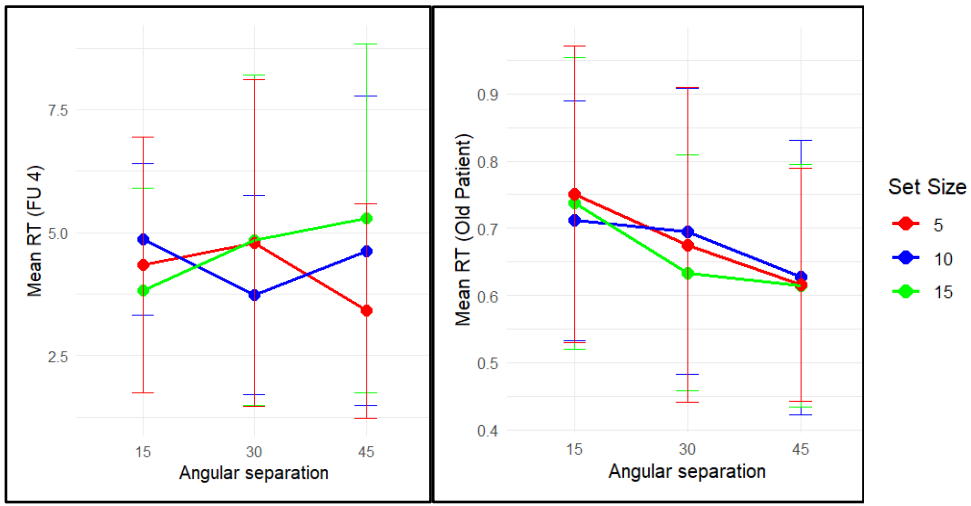

*4.1.2 Accuracy Performance for Human Participants*

Accuracy performance comparison for human groups as affected by endogenous stimuli factors such as set size (Figure 13), relative angle difference between target and distractor (target angle - distractor angle) (Figure 14), and spatial distribution of target relative to its nearest neighbor distractor (Figure 15).

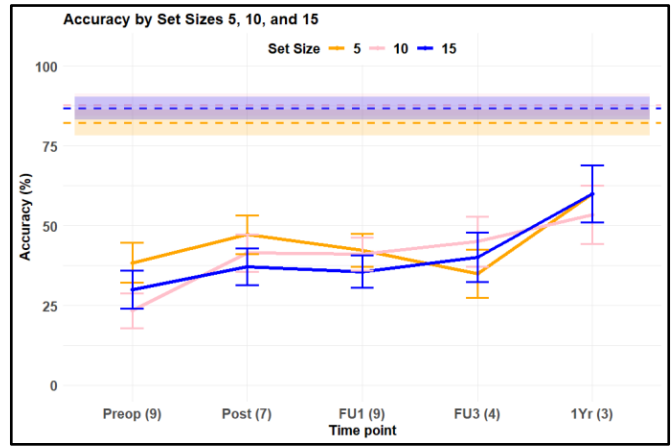

Figure 13: Patient group accuracy grouped by set size. The dashed lines represent performance for blurred control group for the different set sizes.

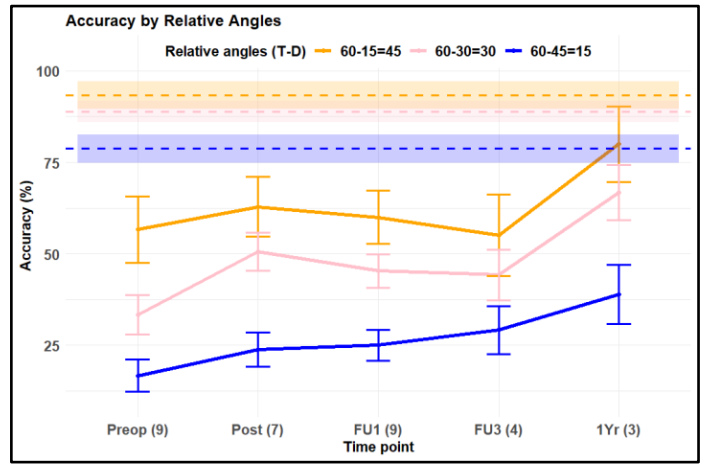

Figure 14: Patient group accuracy grouped by relative angle. The dashed lines represent performance for blurred control group for the different relative angles.

As Figure 14 depicts, there appears a difference in the accuracy based on relative angle difference (which is also called angular separation in model-related terminology). While there appears no difference based on set sizes (Figure 13) indicating towards a parallel search mechanism (as also supported by FIT, as discussed in introduction section).

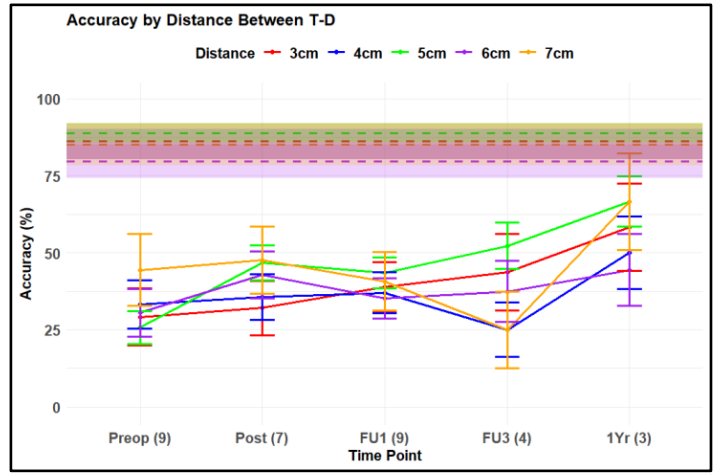

Figure 15: Patient group accuracy grouped by spatial distribution between target and nearest neighbor distractor. The dashed lines represent performance for blurred control group.

## 4.2 Model Performance

### *4.2.1 Replication of original model results*

The model was made to perform the same tasks as provided in the original paper. Figure 16c represents the results reported in the paper. Figure 16d is the result that we got from performing the same task again. It is observed that there is a difference in the performance of these two figures. The results shows that the performance for the homogenous tasks (Figure 16a) is asymmetric in the sense that for target with 20°, model performs search in a serial fashion while for target with 0° angle, it performs search in a parallel manner. In contrast to this, when we performed the task, the results show that for the same combination, the search mechanism is serial for both the cases which also contrasts the human performance pattern. Similar to this task, there are other four tasks that showed similar results for original and replicated experiments are reported in the Supplementary material.

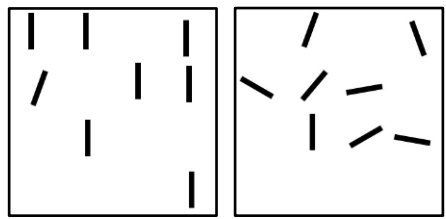

Figure 16: (a) Homogenous task stimuli, (b) Heterogenous task stimuli

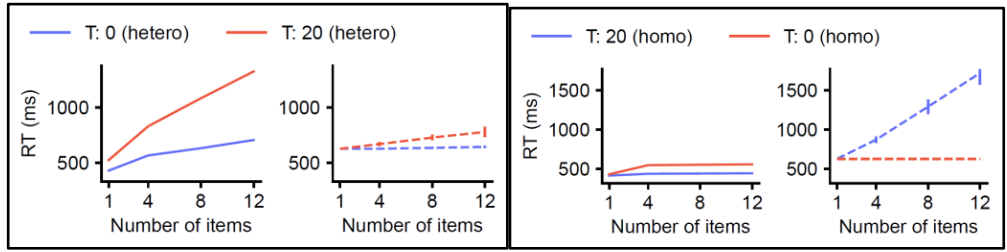

Figure 16: (c) Original results

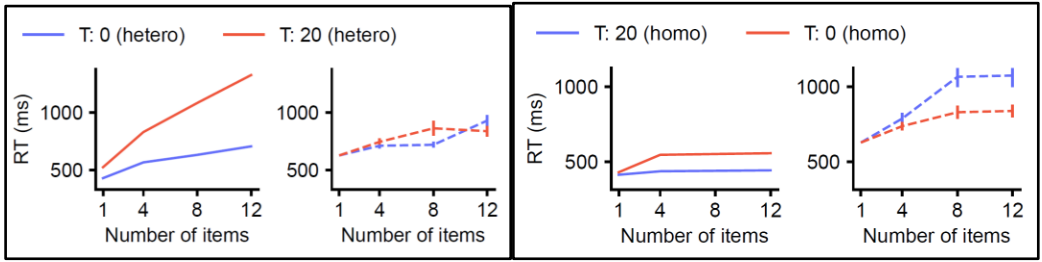

Figure 16: (d) Generated results during replication stage

*4.2.2 Model performance on the modified visual search task*

We observed model performance as determined by the following three endogenous stimuli features: Set size, Angular Separation between Target & Distractor, and Spatial distribution of distractors relative to the target:

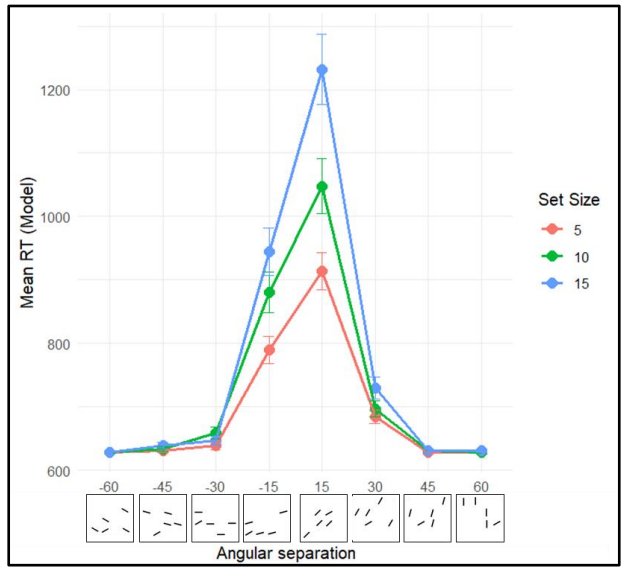

Figure 17: RT performance comparison for the model as affected by stimuli set size and angular separation between target and its nearest neighbor distractor (target - distractor).

The model performance as reported in Figure 17 represents an interesting picture of the mechanism involved in visual search process for the model. This is so because for most angular separation values, there appears no difference between RTs of different set sizes, however for the least angular separation values (namely, 15 degree and negative 15 degree difference), there is a difference between RT based on set sizes, indicating towards the serial search mechanism.

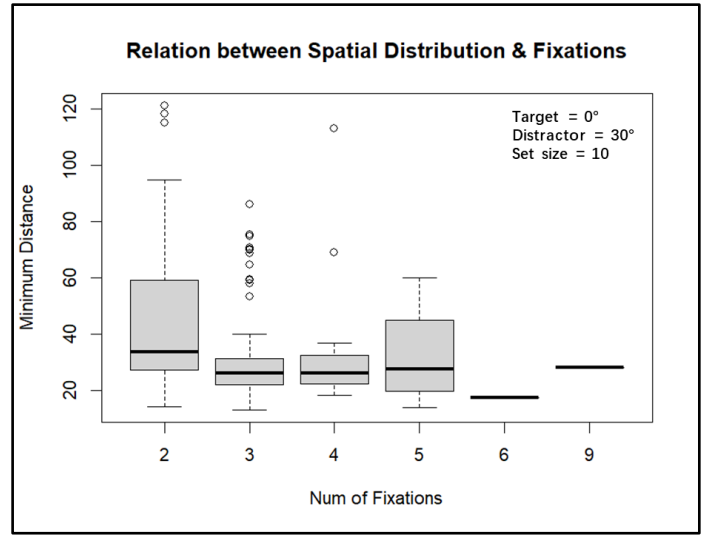

Figure 18: Relationship between spatial distribution of target w.r.t nearest-neighbor distractor and number of fixations required for search process

In order to understand the relationship between spatial distribution of target w.r.t nearest-neighbor distractor and the time taken to respond, we calculated the distance from the target to nearest distractor using image processing tools in python in terms of pixels. There was no grouping in this case (as in human performance case) for the distances. However, similar to the human case (Figure 15), there appears no apparent pattern of change in RT as a function of the distance between target and nearest neighbor distractor (Figure 18). From the perspective of patients' performance, this is an important feature since some of the research by Project Prakash highlights that in their early age of visual skills (after sight-recovery), patients tend to percieve objects as over-fragmented (Ostrovsky et al., 2009). Therefore, in an attempt to simulate patient performance, this distance remains an important stimuli feature.

*4.2.3* **Visualization of model search mechanism**

The visual search mechanism using eccNET model includes reproduction of intermediate attention maps for each item in the stimuli. The attention maps for each fixation appear like heat maps showing the weights of the network's activations toward regions likely containing the target. Moreover, based on the sequence of activation of the various regions of the stimuli, a saccade path map can be created. In the figure below (Figure 19), two cases of set size = 5 and 15 are compared using their attention maps and saccade path map. The target stimuli shown here has target of 60 degrees and distractor at a difference of 15 degrees from the target (i.e., distractor angle is 45 degrees).

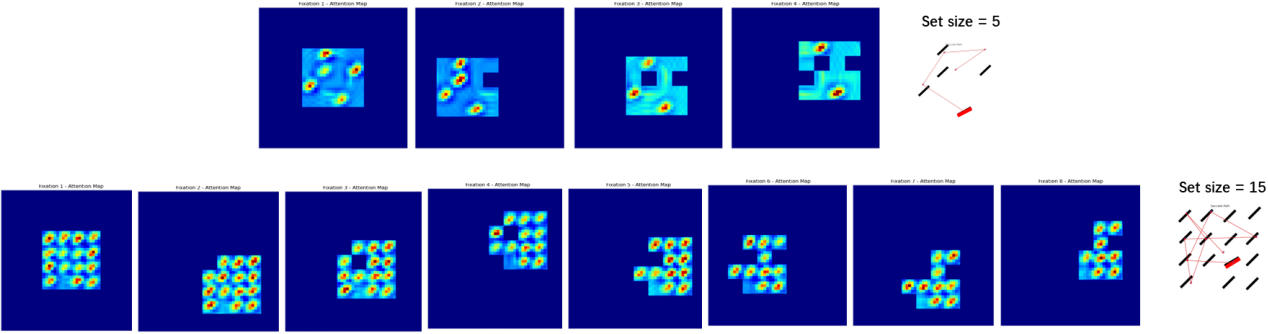

Figure 19: Visual search mechanism visualization: Easy vs Difficult stimuli comparison

Similarly, Figure 20 represents a comparison of visual search mechanism of a stimuli with 15 degree (distractor angle = 45 degrees) difference and a negative 15 degree difference (distractor angle = 75 degrees).

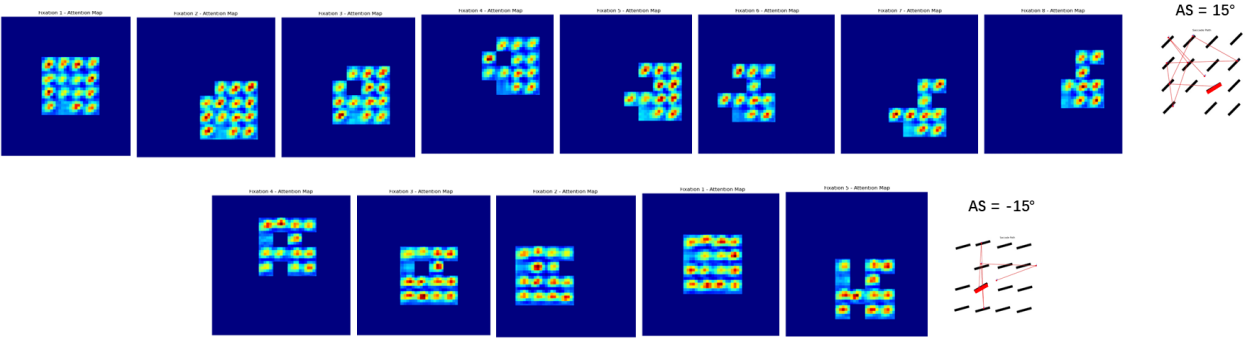

Figure 20: Visual search mechanism visualization: 15-degree vs negative 15-degree difference between target and distractor comparison

## 4.3 Attempts to Simulate Patient Performance using Model

We made several ablations in the model and the task to simulate human-like performance. The ablations were made based on theoretical background based on existing literature.

### *4.3.1 Angular Separation based Task-Ablation:*

Wolfe et al., (1999) suggested that differences of around 15–90° between target and distractors allow for efficient visual search. This principle formed the foundation of our first ablation, wherein various target angles were combined with homogeneous distractors to design different stimuli, enabling comparative analysis between model performance and human performance.

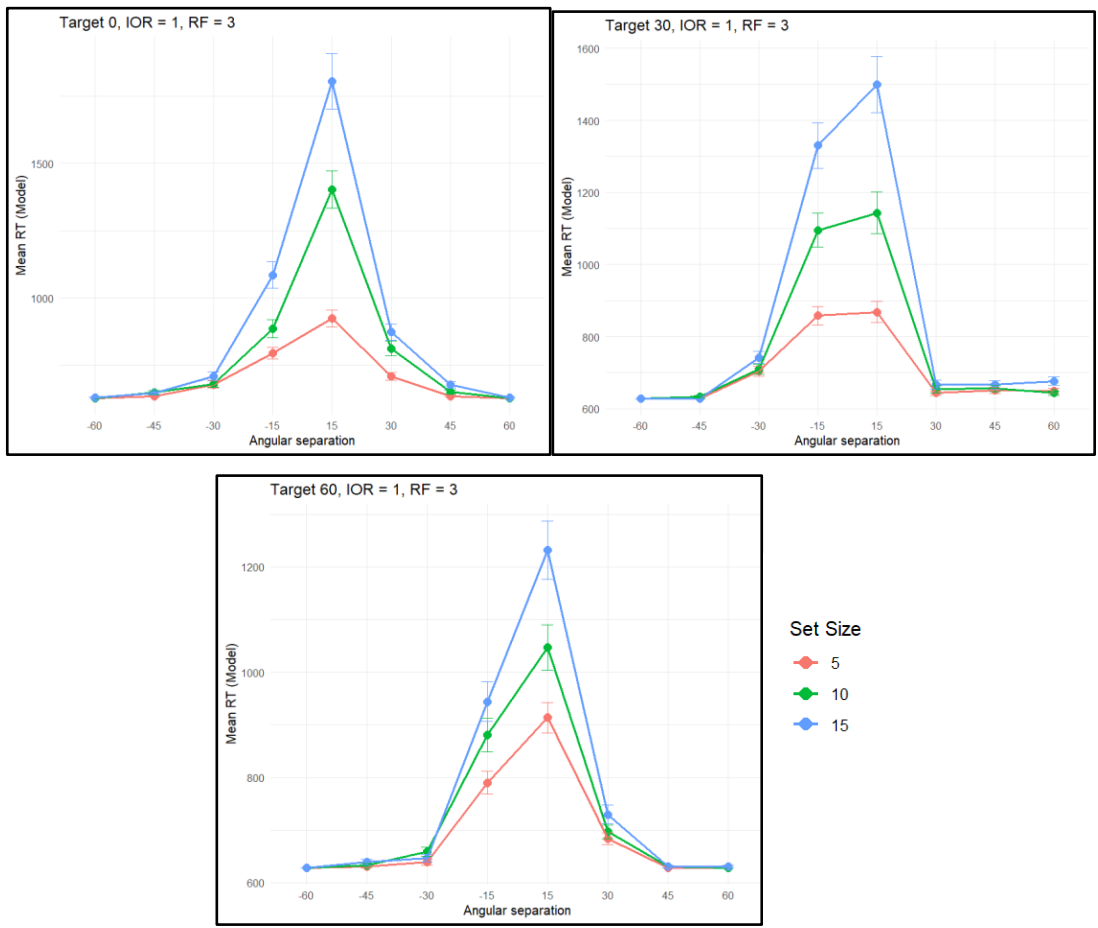

Figure 21: Model performance on visual search task having stimuli with target angles: 0°, 30°, and 60° with homogenous distractors at different angle differences (angular separation).

### *4.3.2 Receptive Field Size based Model-Ablation:*

We also introduced several model ablations to bring the model performance closer to human performance in the different cases under study. The first ablation was performed in terms of receptive field size for the various layers in the model. This was inspired from Ostrovsky et al., (2009) who who documented over-fragmentation in visual parsing among prakash patients during object recognition tasks. We hypothesized that if the model similarly parsed images into smaller fragments at different processing stages, discernible differences in RT performance might emerge. Results from this exploration are documented in the supplementary material, where we modified the receptive field size (originally 2 for all layers) to 1, which produced alterations in RT patterns across various angular separations. In a separate condition, we modified only the fifth layer's RF size to 1 while maintaining original parameters for the remaining four layers, which similarly resulted in subtle RT modifications while preserving the overall pattern observed in the first condition. The principal ablation results, wherein RF size was increased from 2 pixels to 3 pixels, are presented in Figure 22.

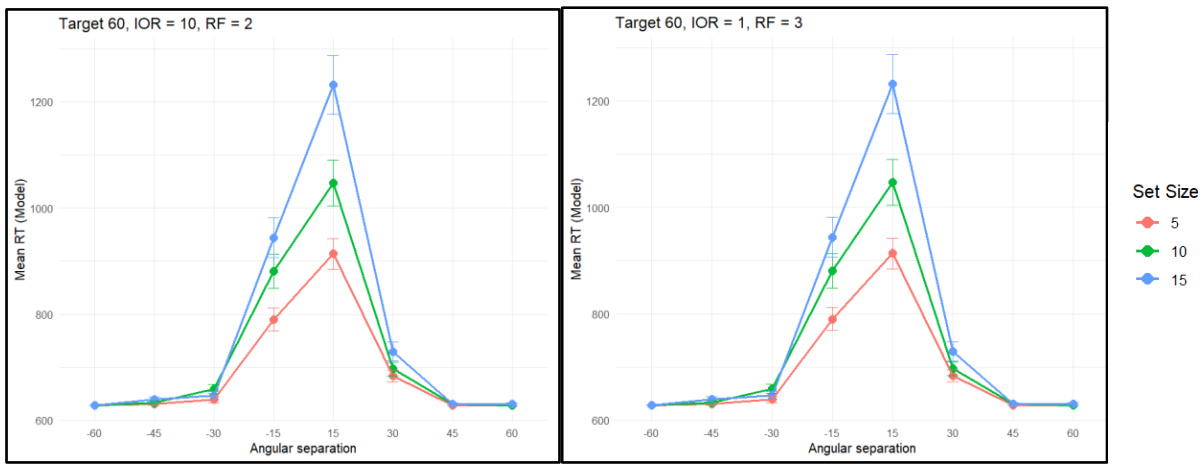

Figure 22: Receptive field size based model ablation results for RF size = 2 and 3 pixels respectively.

*4.3.3 Inhibition of Return based Model-Ablation:*

The next ablation was changing the inhibition of return (IoR) value which was originally 10 by a factor of 1/2 and 1/10. This modification is inspired by findings from Gupta et al., (2022) which documented that immediately following sight-restoring surgery, patients exhibited diminished visual memory capacity despite gaining patterned vision. However, their visual memory performance improved substantially over subsequent months, demonstrating plasticity in visual memory mechanisms even after early visual deprivation. Consequently, we hypothesized that inhibition of return would be reduced in the patient population. Nevertheless, the results observed for IoR modifications (Figure 23) revealed no significant effect on RT performance metrics.

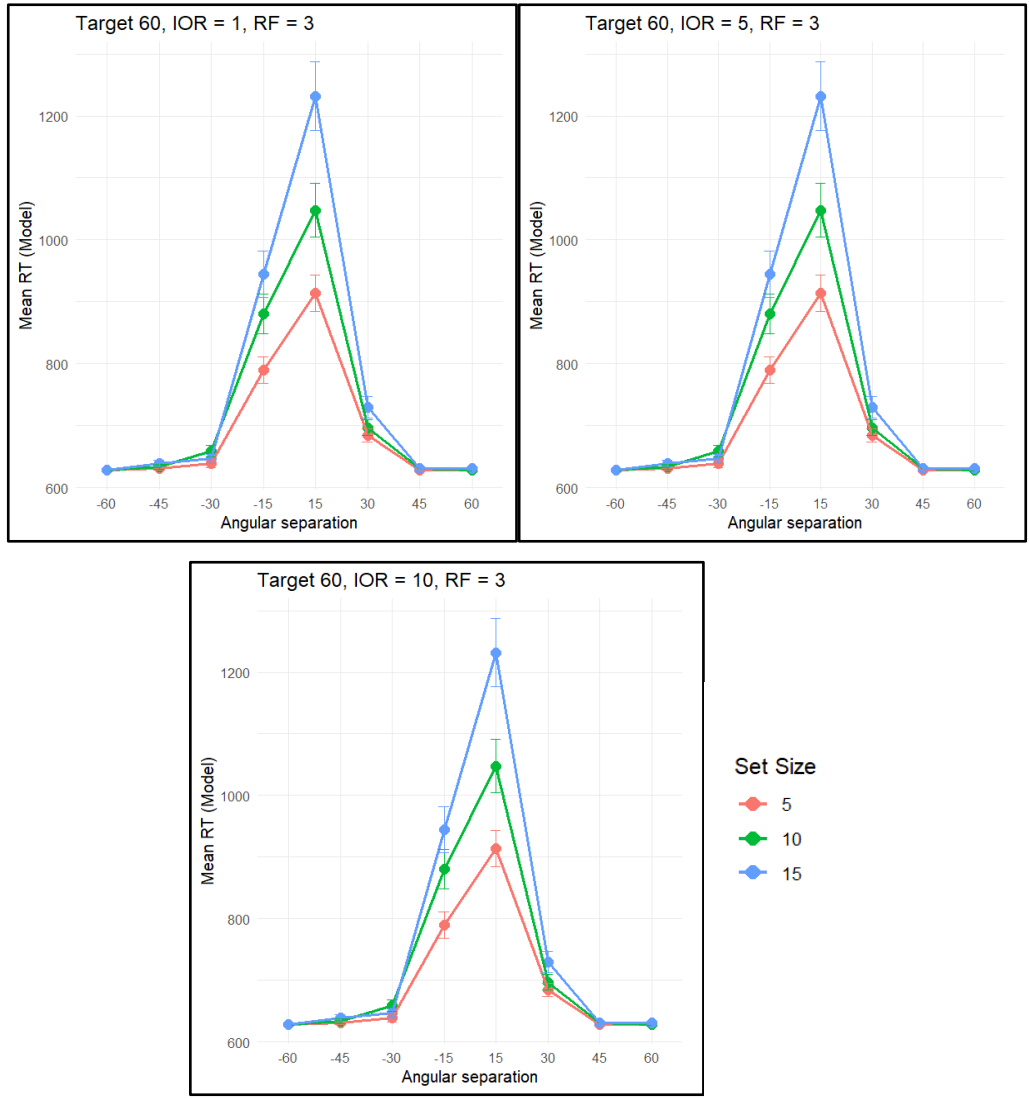

Figure 23: Inhibition of return based model ablation results for IoR size = 1, 5, and 10, respectively.

### *4.3.4 Dataset Batch Size based Model Ablation:*

Additionally, we reduced the dataset batch size to one-third of its original value. This modification was predicated on the concept of gradual learning through developmental visual exposure. In the case of prakash patients, this exposure remained restricted prior to surgical intervention and subsequently increased, albeit at an accelerated rate, following sight-restoration procedures. However, modifications to batch size did not appreciably alter model performance in terms of reaction time (Figure 24).

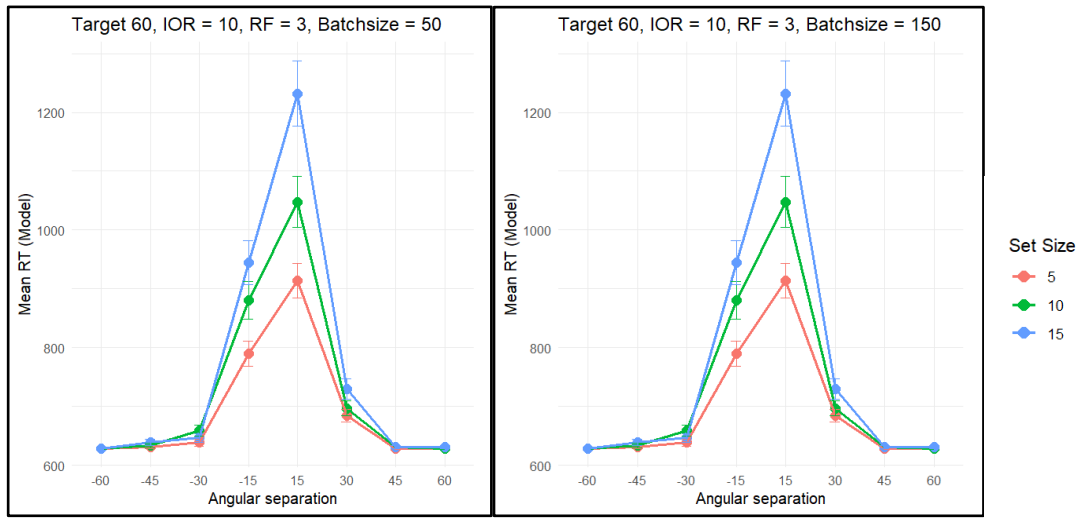

Figure 24: Batch size based model ablation results for Batch size = 50 and 150 respectively.

## 4.4 Discussion

The performance data from human participants demonstrated probable differences between patient and control populations across multiple dimensions of visual search tasks.The most trusted findings are in terms of accuracy measurements, which demonstrated remarkable sensitivity to angular separation between targets and distractors (Figure 14), yet showed puzzling invariance to set size manipulations (Figure 13). This pattern indicates to the parallel search mechanisms in these participants which is also supported by Feature Integration Theory albeit with important caveats specific to this population such as low visual acuity, and other visual impairments.

Patient performance exhibited a graded pattern that varied with angular separation. Despite their recent sight restoration, these individuals apparently developed orientation discrimination mechanisms, although maintaining a reduced accuracy compared to blurred acuity controls. The lack of systematic relationship between target-distractor proximity and performance (Figure 15) runs counter to intuition that stronger interference from nearby distractors must affect performance negatively. This unexpected finding points to altered attentional allocation mechanisms in our patient population, possibly reflecting incomplete development of spatial inhibition processes usually crucial for efficient visual search. In order to check for the effect in typical cases (which is not present in form of control data), we checked for spatial distribution effect for the model performance data (Figure 18). An absence of a clear trend here also does not give clear indication about effect of distractors' closeness to the target. Although, the simplicity of this experimental stimuli must also be considered as a caveat for a definitive conclusion.

From the trends visible in human performance data (in terms of accuracy), we deduced the idea that various target angles and distractors combination must be explored to understand the effect of target orientation and target-distractor angular separation through the model performance. However, a troubling discrepancy emerged between original computational model results and our replication attempts. While the original model showed asymmetric search patterns for homogeneous stimuli: parallel search for 0° targets versus serial search for 20° targets, our replication consistently yielded serial search patterns in both conditions. This inconsistency raises methodological questions regarding reproducibility in computational vision research since no parameters were changed for the process, due to which the discrepancy in the two results could not be definitely explained. We can only speculate that the original implementation likely contained unspecified parameter initializations that produced different outcomes or some subtle implementation details in terms of machinery requirement could dramatically impact model behavior.

Our attempts to simulate patient performance through various model ablations produced a mixed bag of results. The angular separation manipulations (Figure 21) showed that the model, like human participants, exhibits sensitivity to orientation differences between targets and distractors. This suggests that orientation-based search mechanisms might be reasonably well-captured by current computational frameworks.

Receptive field size manipulations (Figure 22) showed that there was no change in model performance if it is increased however when it is decreased to a point where it becomes computationally non-plausible (i.e., receptive field size = 1 pixel), then it changes the performance for when target is 60 degrees (Supplementary figure 3). There is no clear conclusion derived from this ablation regarding human-like performance simulation. Given well-documented visual memory deficits in newly-sighted individuals (Gupta et al., 2022), we had hypothesized these would manifest as reduced inhibition of return in model performance. The model's insensitivity to these parameter changes raises questions: Either the model inadequately implements inhibitory mechanisms, or inhibition of return plays a less significant role in the visual memory deficits of this population than we initially thought.

Similarly, batch size manipulations (Figure 24) intended to mimic gradual learning processes barely affected performance. Simple quantitative reductions in training exposure evidently cannot capture the qualitative learning differences that characterize our patient population. This suggests we need more sophisticated developmental models, those that can incorporate critical period constraints and compensatory plasticity mechanisms in order to faithfully simulate these patients' learning trajectories.

# CHAPTER 5: CONCLUSION AND SCOPE OF FURTHER WORK

Our investigation into visual search mechanisms among sight-recovered individuals yielded several noteworthy findings while simultaneously opening opportunities for further exploration. The computational modeling approach, despite its limitations, provided valuable insights into the perceptual mechanisms that may underlie the performance patterns observed in Project Prakash patients.

A critical direction for extending this work involves addressing the fundamental task differences between human participants and computational models. Humans were instructed to identify the "odd one out" bar, whereas the model was programmed to search for a specified target. This distinction reflects the requirement of substantially different cognitive resources and processes. The human task necessitates an active comparison process utilizing visual working memory; participants must establish a perceptual norm from the display and identify deviations from this emergent template (as also described in guided search model 6.0). In contrast, the model performs what amounts to template matching against a predetermined target specification.

It is necessary to examine this disparity using different paradigms. A suitable foundation is provided by the picture visual search task for which we have data from the Prakash patient group (presented in Supplementary Figure 4). A number of methodological changes would be necessary to put this paradigm into practice. Our existing modeling framework is unable to handle the color and greater dimensions of the original stimulus used with human participants. We pre-processed all stimuli by separating target elements from their contexts and converted them to grayscale as a temporary fix. The original stimulus (left) and our manually made ground truth mask (right), which indicates the target location (in this case, an apple shape), are both displayed in Figure 25 to demonstrate this method.

Beyond task considerations, our modeling efforts demonstrated important shortcomings in existing computational frameworks. Although our ablation experiments focused on different visual processes that may be affected in the patient population, these changes did not function together. Actual visual impairments most likely result from the interplay of several faulty processes. Combinatorial parameter modifications must be used in future research to capture these interaction effects.

Another drawback of our modeling technique is that it is primarily phenomenological. Although these models offer useful explanations of performance, they are devoid of neurobiological limitations that could more accurately depict the developmental quirks of sight-recovered people. Our understanding of visual recovery would be significantly improved by more complex models that incorporate the neurophysiological characteristics of the growing visual system, especially with regard to experience-dependent plasticity processes and critical period limitations.

A step toward comprehending visual perception after prolonged deprivation is represented by the current work. Through the use of computational modeling techniques in conjunction with psychophysical measures, we have started to pinpoint both maintained and modified pathways in this distinct group. Building on these findings, future study could improve theoretical knowledge of visual development and guide rehabilitation plans for patients undergoing sight-restoring procedures.

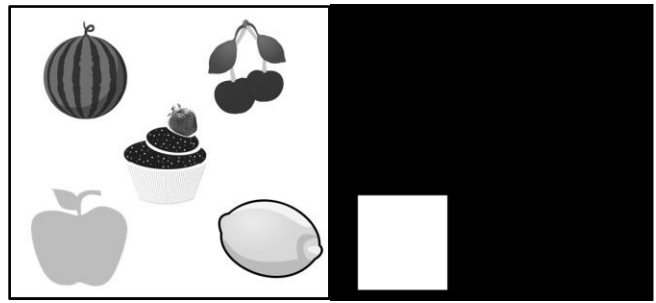

Figure 25: Image visual search task stimuli (left) and its corresponding ground truth mask, considering the apple shape is the target

**SUPPLEMENTARY MATERIAL**

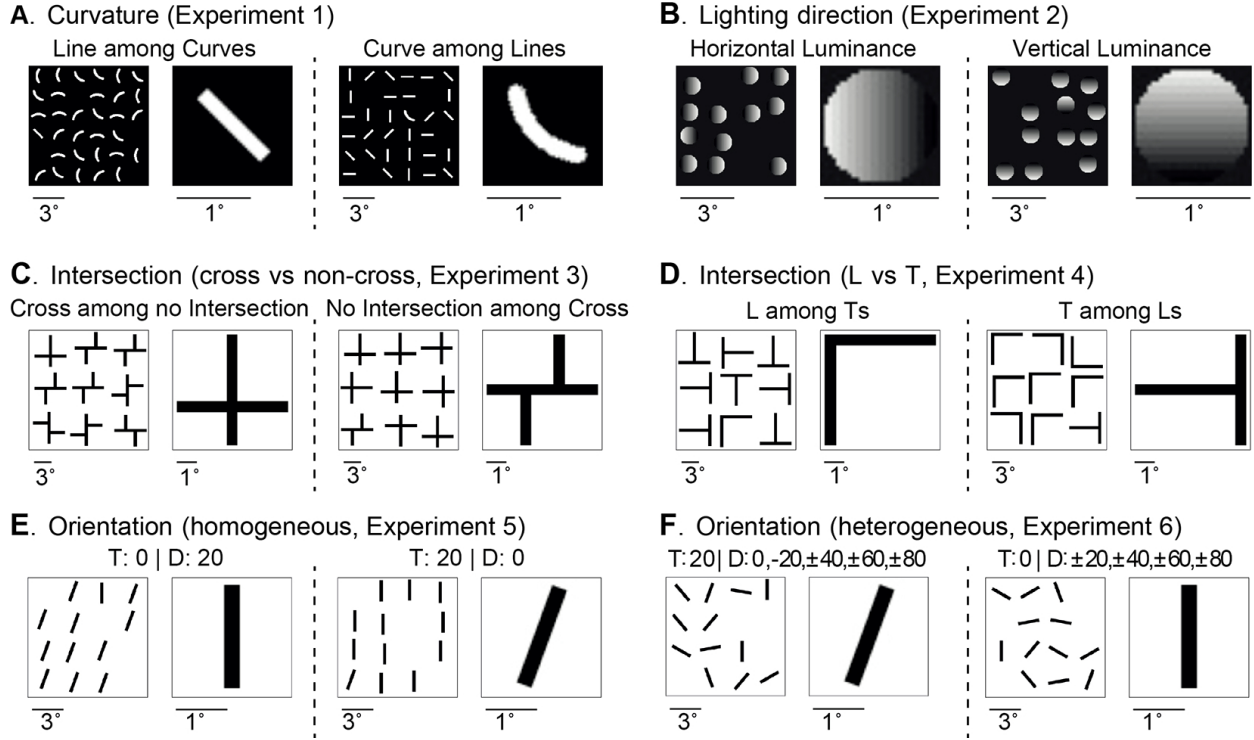

Supplementary Figure 1. Experiment 1. Searching for a curve among lines and vice versa ([48]). B. Experiment 2. Searching for vertical luminance changes among horizontal luminance changes and vice versa ([22]). C. Experiment 3. Searching for shapes with no intersection among crosses and vice versa ([46]). D. Experiment 4. Searching for rotated Ts among rotated Ls and vice versa ([46]). E. Experiment 5. Searching for oblique lines with fixed angles among vertical lines and vice versa ([47]). F. Experiment 6. Similar to Experiment 5 but using oblique lines of different orientations ([47]). In all cases, subjects find the target faster in the condition on the right.

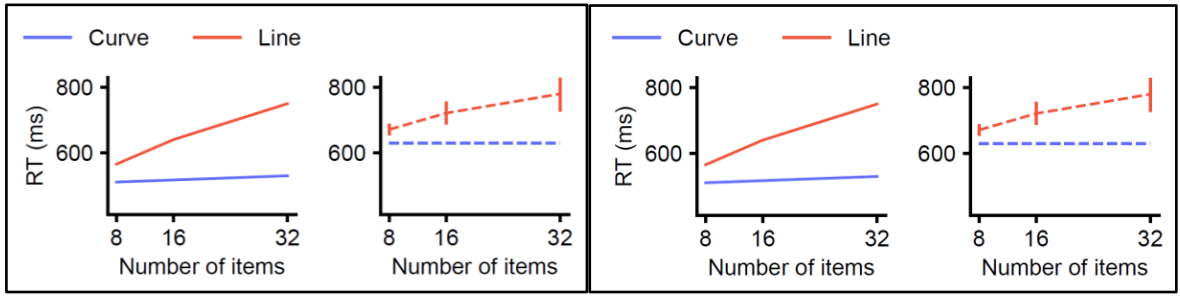
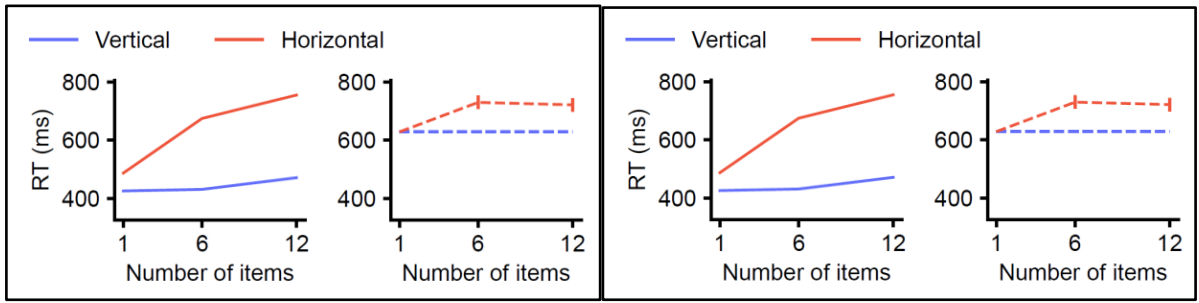
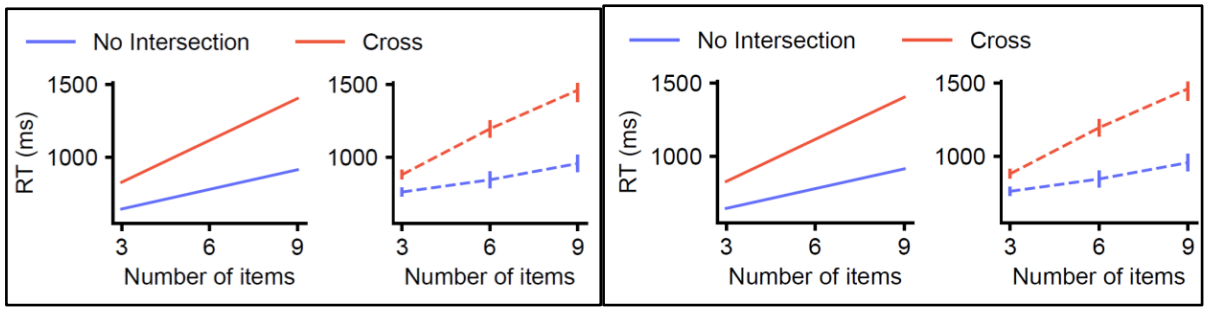
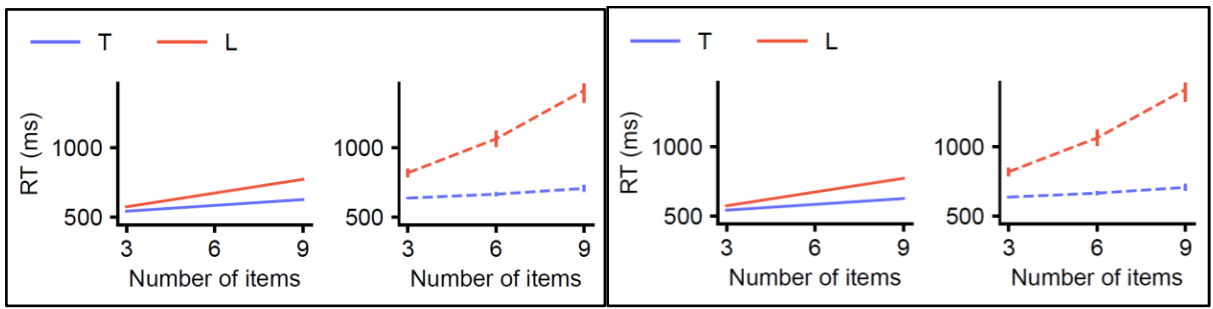

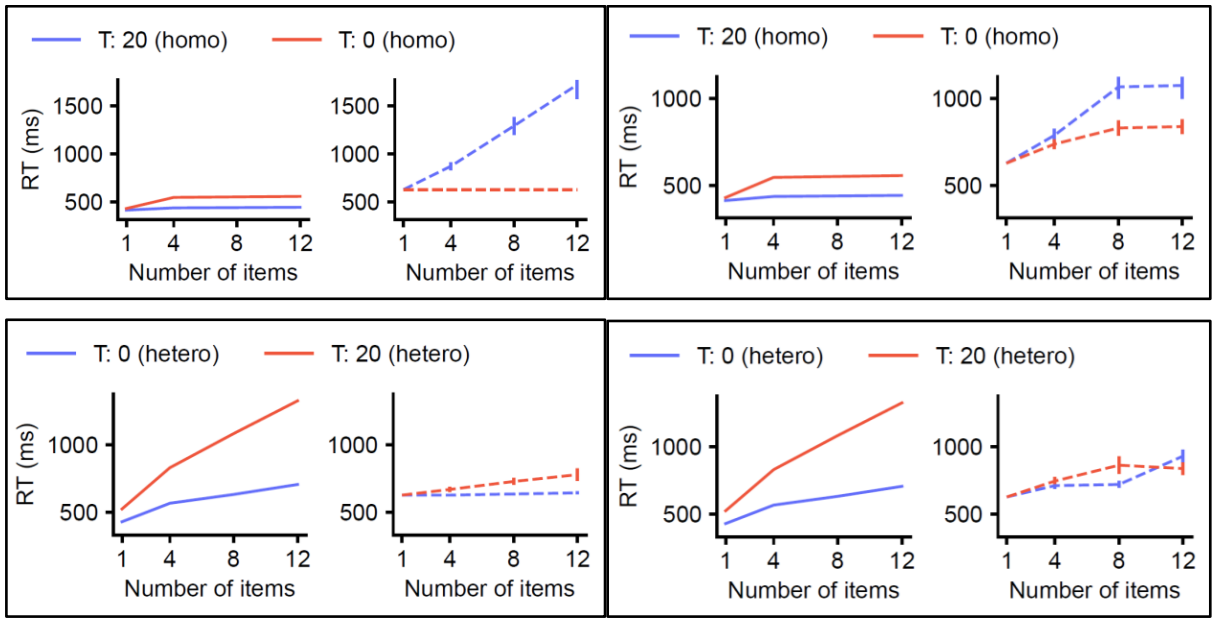

Supplementary Figure 2. Original output for each experiment, as given in the paper compared to the replicated output for the same task.

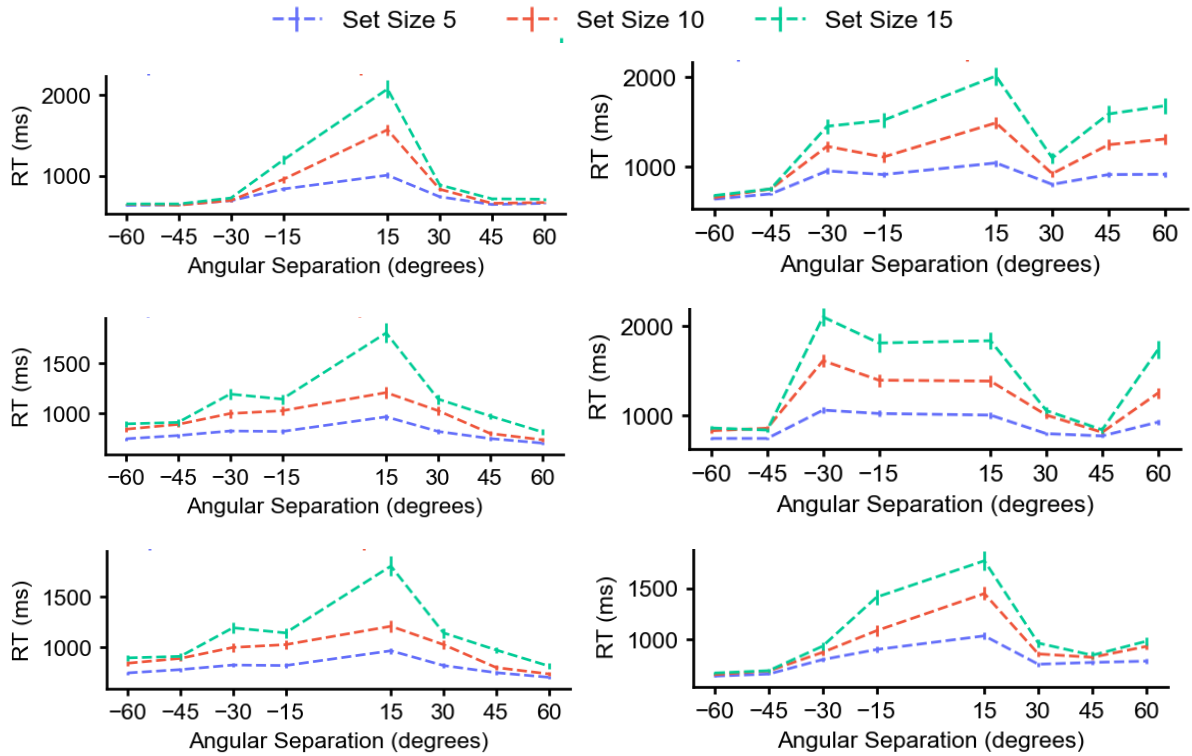

Supplementary Figure 3. Receptive field size is modified in three different ways for the same set of stimuli with target angles = 0 degrees (left) and 60 degrees (right). RF size ablations are as follows in the sequence of figures: 2 (all layers), 1 (all layers) and 1 (4 layers) + 2 (5th layer)

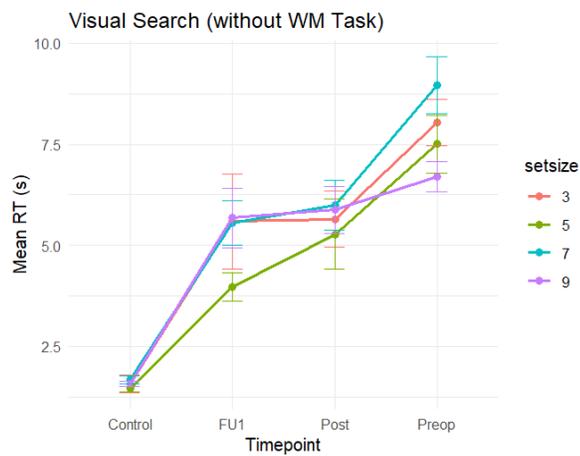

Supplementary Figure 4: Results for human participants groups on the image visual search task

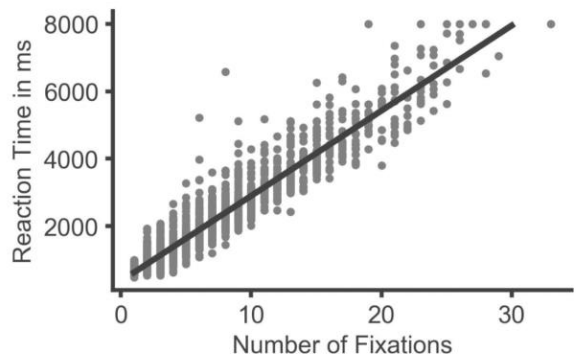

Supplementary Figure 5: Reaction time grows linearly with the number of fixations. Each gray point represents a trial. A line was fit to these data: R(ms) = α ∗ n + β. A fit using linear least

square regression gave α = 252.359 ms/fixation and β = 376.271 ms ($r^2$ = 0.90, p << 0.001). This linear fit was used throughout the thesis to convert the number of fixations in the model to reaction time values in milliseconds for comparison with human data